\documentclass[twocolumn,aps,superscriptaddress,prl,longbibliography]{revtex4-1}
\usepackage{palatino}
\usepackage[latin1]{inputenc}
\usepackage{epsf}
\usepackage{amsmath,amssymb}
\usepackage{latexsym}
\usepackage{calc}
\usepackage{xcolor}

\usepackage{graphicx}
\PassOptionsToPackage{hyphens}{url}\usepackage{hyperref}
\newcommand{\ben}{\begin{equation}}
\newcommand{\een}{\end{equation}}
\newcommand{\bea}{\begin{eqnarray}}
\newcommand{\eea}{\end{eqnarray}}

\def\sss{\scriptscriptstyle\rm}

\def\1s{_{1,\sss S}}
\def\2s{_{2,\sss S}}

\def\eps{{\epsilon}}

\begin{document}
\title{Exact-factorization-based surface-hopping without velocity adjustment}
\author{Lucien Dupuy}
\affiliation{Department of Physics, Rutgers University, Newark 07102, New Jersey USA}
\author{Anton Rikus}
\affiliation{{University of M\"unster, Organisch-Chemisches Institut and Center for Multiscale Theory and Computation, M\"unster, Germany}}
\affiliation{Department of Physics, Rutgers University, Newark 07102, New Jersey USA}
\author{Neepa T. Maitra}
\affiliation{Department of Physics, Rutgers University, Newark 07102, New Jersey USA}

\date{\today}

\begin{abstract}
While surface-hopping has emerged as a powerful method to simulate non-adiabatic dynamics in large molecules, the {\it ad hoc} nature of the necessary velocity adjustments and decoherence corrections in the algorithm somewhat reduces its reliability. Here we propose a new scheme that eliminates these  aspects,  by combining the nuclear equation from the quantum trajectory surface-hopping approach with the electronic equation derived from the exact factorization approach. The resulting method, denoted QTSH-XF, places surface-hopping on a firmer ground and is shown to successfully capture dynamics in Tully models and in a linear vibronic coupling model of the photo-excited uracil cation. 
\end{abstract}

\maketitle
The highly correlated motion of electrons and ions when a molecule is driven away from equilibrium gives rise to a rich range of phenomena, occurring both naturally (e.g. in photosynthesis\cite{TMF11,CTBST17}, vision\cite{Cerullo_N2010,MXHG16}) as well as in technological applications (e.g. photocatalyst design\cite{YLC21,AVTP22}, molecular motors\cite{FPMC19,   WMG23}). Theoretical simulation of these processes reveals the fundamental nature of coupled electron-ion dynamics in general while aiding our understanding of the mechanisms of the specific processes involved, and predictions of new phenomena.

Since its inception about fifty years ago, trajectory surface-hopping (SH) has remained one of the most the popular methods to simulate such coupled electron-ion dynamics~\cite{TP71,T90,T98,WAP16,SJLP16,MMG20}. Due to its computational efficiency, it can treat complex systems with a large number of degrees of freedom  without needing to make {\it a priori} assumptions about the relevant configurations encountered in the dynamics.  SH couples a classical treatment of nuclear dynamics with a quantum treatment of electrons, but nevertheless captures the quantum phenomenon of nuclear wavepacket-splitting, which is lacking in another widely-used and practically-efficient method, Ehrenfest dynamics. These features are enabled by having independent trajectories each running on a single adiabatic electronic potential energy surface (PES) at a given time, making hops between them according to a stochastic algorithm that depends on the trajectory's velocity, non-adiabatic coupling vectors  (NACV), and the electronic coefficients that are evolved quantum-mechanically  self-consistently with the classical nuclear trajectory's position.

However, SH has proven difficult to derive from first-principles~\cite{SOL13,LZ18}, and it requires at least three {\it ad hoc} procedures in order to give physically sound results. First, after a trajectory hop from one PES to another, the nuclear velocity is re-scaled so as to preserve the total energy of the nuclear trajectory, and there is no unique way to do this,  e.g. isotropically, or along the NACV; results differ depending on how it is done~\cite{CGB17,B21,TSF21,VIHMCM21,MR23}. Moreover, the imposition of energy conservation at an individual trajectory level is too tight a constraint: physically, it should be the energy of the ensemble of trajectories that is conserved, since it is the energy of the quantum wavepacket as a whole,  whose density is mimicked by the trajectory ensemble, that is conserved. 
Further, if a hop is rejected because the trajectory does not have enough energy, channels are closed which would be open in a quantum treatment.  A second {\it ad hoc} aspect is that, for such a frustrated hop, a choice must be made as to whether the trajectory's velocity is either kept, or always reversed, or reversed under some conditions~\cite{CGB17,HT94,JT03,PMFGDG19}. 
Third, although the fewest-switches hopping probability was designed such that the fraction of trajectories evolving on a given PES agrees with the trajectory-average of the modulus-square of the electronic coefficient for that state, the latter is not accessible due to the independent-trajectory framework of SH; the deviation in these two quantities is referred to as internal inconsistency. The problem is often phrased as ``overcoherence" since the electronic coefficients remain always evolving coherently, while the trajectory they are associated with is always collapsed to a single PES.  While several decoherence corrections are routinely applied they are largely {\it ad hoc}, even if physically motivated, and not always reliable. 

Modified SH methods have been developed that overcome some of these three problems. In quantum trajectory SH (QTSH)~\cite{M19b,M20} which is an independent-trajectory approximation of consensus SH~\cite{M16b,M16cc}, a ``quantum force"  arises in the nuclear equation that eliminates the velocity rescaling procedure while ensuring energy conservation over the ensemble of independent trajectories in cases where the internal consistency is not broken. The overcoherence problem remains however; a correction was proposed outside the independent-trajectory framework. 
On the other hand, SH based on the exact factorization approach (SHXF)~\cite{HLM18,HM22,PyUNIxMD} incorporates a first-principles description of decoherence, through an electronic equation derived from the exact factorization equations~\cite{AMG10,AMG12,MAG15}, but the velocity adjustment problems remain. 

In this work, we combine the nuclear equation from QTSH with the electronic equation from SHXF, resulting in an independent-trajectory SH scheme, QTSH-XF, that eliminates all of the three {\it ad hoc} aspects above, thus providing a more robust method than (decoherence-corrected) SH, QTSH, and SHXF, while retaining the practical efficiency of independent-trajectory methods. We demonstrate its performance on Tully's extended coupling region (ECR) model~\cite{T90} and on a linear vibronic coupling model of the uracil cation~\cite{AKM15}. 

In standard fewest-switches SH~\cite{T90,T98}, an ensemble of independent nuclear trajectories $\{\textbf{R}^{(I)(t)}\}$  is evolved, each associated with an electronic wavefunction whose expansion coefficients in the Born-Oppenheimer (BO) basis evolve as
\ben \label{eq:SHelectronic}
\begin{split}
    &\dot C_l^{(I)} = -\frac{i}{\hbar}\epsilon^{\mathrm{BO}}_l(\textbf{R}^{(I)}) C_l^{(I)} - \sum_k C_k^{(I)} \sum_{\nu} \dot{\textbf{R}}_\nu^{(I)}\cdot\textbf{d}_{\nu,lk}^{(I)},
\end{split}
\een
where $\textbf{d}_{\nu,lk} = \langle \Phi_{l}\vert\nabla_\nu\Phi_k\rangle$ is the NACV, while
the equations of motion for the nuclei are
\ben\label{eq:SHnuclear}
    \dot{\textbf{R}}^{(I)}_\nu = \frac{\textbf{P}^{(I)}_\nu}{M_\nu} \quad \text{ and } \quad\dot{\textbf{P}}_\nu^{(I)} = - \nabla_\nu \epsilon^\mathrm{BO}_a(\textbf{R}^{(I)}),
\een
where $\epsilon^\mathrm{BO}_a(\textbf{R}^{(I)})$ denotes the  PES that trajectory ${(I)}$ is evolving on (i.e. the active surface), evaluated at its current geometry $\textbf{R}^{(I)} = \textbf{R}^{(I)}(t)$ and $\nu$ labels the nucleus. 
The fraction of trajectories that are evolving on the $l$th surface at a given time $t$, $\Pi_l(t) = \sum_I^{N_{\rm traj}}N^{(I)}_{l}(t)/N_{\rm traj}$, defines an electronic population distinct from the population obtained directly from the electronic equation, $\rho_{ll}(t) =  \sum_I^{N_{\rm traj}}\rho^{(I)}_{ll}(t)/N_{\rm traj}$, with  $\rho^{(I)}_{ll}(t)=\vert C^{(I)}_l(t) \vert^2$, and usually it is $\Pi_l(t)$ that is ultimately recorded as the electronic population.  
The fewest-switches hopping probability was derived to try to minimize the difference between the two population measures (i.e. the ``internal inconsistency")~\cite{T90}: $P_{a\rightarrow k}^{(I)}$ for an instantaneous hop of trajectory ${(I)}$ from active surface $a$ to $k$ is calculated from 
\ben\label{eq:SHhoppingprobability}
    P_{a\rightarrow k}^{(I)} = {\rm max} \left\{0,-\frac{\displaystyle 2 \text{Re}\{C^{(I)*}_a C^{(I)}_k\} \sum_\nu \textbf{d}_{\nu,lk}^{(I)}\cdot \dot{\textbf{R}}^{(I)}_\nu}{|C^{(I)}_a|^2} \Delta t \right\}.
\een
Then, at each time step, a hop from the active state $a$ to state
$n$ is made if $\sum_n^{k-1} P_{a\rightarrow k}^{(I)} < r <  \sum_n^{k} P_{a\rightarrow k}^{(I)}$, where $r$ is a random number
uniformly distributed in $[0,1]$, provided the change in potential energy after the hop does not exceed the kinetic energy before the hop. If the kinetic energy is exceeded, the hop is rejected (also known as frustrated, or forbidden), and a choice is made as to whether the trajectory's velocity is reversed or kept. When a hop is allowed, the velocity after the hop is adjusted so as to ensure energy conservation, and again a choice is made as to how this is done. 
Two common choices are to scale the velocity along the direction of the NACV, or to scale isotropically~\cite{HT94,PMFGDG19,CGB17,JT03,B21,VIHMCM21}; while the former has been justified by semiclassical arguments, it leads to more frustrated hops so a greater violation of internal consistency, but the latter is unphysically size-extensive. 

The disconnect between the coherent electronic evolution and the stochastic single-surface evolution of the nuclear trajectories is partially patched through the use of decoherence corrections, e.g.~\cite{GP07,GPZ2010,SJLP16,SOL13,JFP12,ZNJT04,LAP16,VIHMCM21,MMG20}, which, like the SH-scheme itself, tend to be somewhat {\it ad hoc}, even if physically motivated. Relatively recently, SHXF was proposed~\cite{HLM18,HM22,PyUNIxMD}, in which decoherence arises naturally from a term {\it derived} from the exact factorization approach. 
In SHXF~\cite{HLM18,HM22,PyUNIxMD}, while the nuclear trajectories follow the usual SH algorithm, the electronic equation of motion is rigorously derived from a mixed quantum-classical treatment~\cite{MAG15,AMAG16,AG21} of the exact factorization approach~\cite{AMG10,AMG12,VAM22} and has the following form
\ben \label{eq:SHXFelectronic}
    \dot C_l^{(I)} = \dot C_{l,{\rm SH}}^{(I)}
    - \sum_{\nu} \frac{\mathcal{Q}_\nu^{(I)}}{\hbar M_\nu} \left[ \sum_k |C_k^{(I)}|^2 \textbf{f}_{k,\nu}^{(I)} - \textbf{f}_{l,\nu}^{(I)}  \right] C_l^{(I)} \;,
\een
where $\dot C_{l,{\rm SH}}^{(I)}$ is the SH evolution from Eq.~(\ref{eq:SHelectronic}), 
$\textbf{f}_{l,\nu}^{(I)}$ is the time-integrated force on PES $l$, and $\cal{Q}$ is the nuclear quantum momentum:
\ben
\textbf{f}_{l,\nu}^{(I)}(t) = - \int^t \nabla_\nu \epsilon_\mathrm{BO}^{l,(I)}(t') \mathrm{d}t' \;,\; {\cal Q}_\nu^{(I)} = -\left.\frac{\nabla_\nu \vert\chi\vert}{\vert\chi\vert}\right\vert_{{\textbf{R}^{(I)}}}
\een
 Eq.~\eqref{eq:SHXFelectronic} correlates the electronic evolution with the distribution of nuclear trajectories via the quantum momentum. To enable the independent-trajectory nature of the scheme, $\cal{Q}_\nu$ is computed with auxiliary trajectories launched on PESs in which the electronic populations have become non-negligible. In addition to describing decoherence from first principles, SHXF describes quantum-momentum-driven transitions that were shown to be essential in multi-state problems such as arise with three-state intersections, and are completely missed by all other SH methods~\cite{VMM22}. 

However, since the nuclear evolution in SHXF is simply taken from that in SH, SHXF inherits the  momentum-adjustment in the hopping events from the nuclear treatment. This not only adds uncertainty to the predictions as discussed earlier, but is also unsettling from a phase-space  
analysis of a quantum-classical Liouville equation derivation of FSSH which shows that the hops should occur locally in phase-space, i.e. the nuclear trajectories should have the same position and momenta after the hop~\cite{M16b}. With a trajectory ensemble representing the phase-space density instead, QTSH derives its evolution equations from a semiclassical limit of the Liouville equation, before taking an independent trajectory approximation~\cite{M16b,M16cc,M19,M19b,M20}. The energy of an individual trajectory is not conserved, but energy conservation over the whole ensemble of trajectories holds up to the deviation in internal consistency. The QTSH equations are:
\begin{subequations}
\begin{align} \label{eq:QTSHRdot}
    \dot{\textbf{R}}^{(I)}_\nu =& \frac{\textbf{P}^{(I)}_\nu}{M_\nu} - \frac{2\hbar}{M_\nu} \sum_{i<j} \text{Im}\big\{\rho_{ij}^{(I)}\big\} \textbf{d}^{(I)}_{\nu,ij} \\
    \begin{split}
    \dot{\textbf{P}}_\nu^{(I)} =& - \nabla_\nu \epsilon^\mathrm{BO}_a(\textbf{R}^{(I)}) \\
    &+ \sum_\mu\sum_{i<j} \frac{2\hbar}{M_\mu} \text{Im}\{\rho_{ij}^{(I)}\}  (\textbf{P}_\mu^{(I)} \cdot \nabla_\mu )  \textbf{d}_{\nu,ij}^{(I)} \label{eq:QTSHRdot2}
    \end{split}
\end{align}
\end{subequations}
where $\rho^{(I)}_{ij}=C^{(I)}_i \big(C^{(I)}_j\big)^*$.
These equations define the following nuclear force
\begin{equation} \label{eq:FQTSH}
    \begin{split}
        M_\nu \ddot{\textbf{R}}^{(I)}_{\nu, {\rm QTSH}} = & -\nabla_\nu \epsilon^\mathrm{BO}_a(\textbf{R}^{(I)})
        + 2 \sum_{i<j}\Delta \epsilon^{\rm BO}_{ij} \text{Re}\{\rho_{ij}^{(I)}\} \textbf{d}^{(I)}_{\nu,ij}\\     
+ 2 \hbar \sum_{i<j}  \textbf{d}^{(I)}_{\nu,ij} &\sum_{l}\sum_{\mu} \frac{\textbf{P}^{(I)}_\mu}{M_\mu} \cdot \text{Im}\left\{\textbf{d}_{\mu,jl}^{(I)} \rho_{il}^{(I)} -\textbf{d}_{\mu,il}^{(I)}\rho_{jl}^{(I)} \right\}
    \end{split}
\end{equation}
where $\Delta\epsilon_{ij}^{\rm BO} = \epsilon^{\rm BO}_i - \epsilon^{\rm BO}_j$, and where terms of order $\hbar^2/M_\nu$ are neglected. We note that the last term in Eq.~(\ref{eq:FQTSH}) is active when a state is coupled to more than one state at a given position through their NACVs; the term gives a generally much smaller contribution to the force compared with the others.
Compared to the SH force, the additional ``quantum forces" in Eq.~(\ref{eq:FQTSH}) allow QTSH to eliminate both the need for momentum rescaling and the rejection of frustrated hops, and energy is conserved over the trajectory ensemble provided it is internally consistent~\cite{M19b,M20,M16cc}.
The electronic evolution as well as the hopping probability are obtained as in fewest-switches SH, but with the velocity $\dot{\textbf{R}}_\nu^{(I)}$ in Eq.~(\ref{eq:SHelectronic}) and (\ref{eq:SHhoppingprobability}) replaced by $\textbf{P}_\nu^{(I)}/M_\nu$. 
The independent-trajectory nature of QTSH  reintroduces the internal inconsistency/overcoherence problem absent in its
``parent" method, consensus SH; a decoherence correction can be applied artificially~\cite{M19b}. 

Since the nuclear equation of QTSH overcomes the {\it ad hoc} velocity adjustment and frustrated-hop problems of traditional SH, and the electronic equation of SHXF overcomes the {\it ad hoc} nature of the decoherence corrections that are applied, a natural development is to combine the electronic equation of SHXF with the nuclear equation of QTSH. This defines the QTSH-XF algorithm, in which electronic coefficients obey Eq. \eqref{eq:SHXFelectronic} while Eqs. \eqref{eq:QTSHRdot} and \eqref{eq:QTSHRdot2} guide the underlying trajectories. The resulting nuclear force obtained by combining those equations and neglecting terms of order $\hbar^2/M_{\nu}$ reads: 
\begin{equation} \label{eq:FQTSHXF}
    \begin{split}
        &M_\nu \ddot{\textbf{R}}_{\nu, {\rm QTSH-XF}}^{(I)}  =  M_\nu \ddot{\textbf{R}}_{\nu, {\rm QTSH}}^{(I)} + \mathcal{F}_Q, \quad \text{where  } \;\mathcal{F}_Q \;\text{  is }
        \\
  & 4 \sum_{i<k}  \text{Im}\{\rho_{ik}^{(I)}\} \textbf{d}_{\nu,ik}^{(I)} \sum_{\mu} \frac{\textbf{Q}^{(I)}_{\mu}}{M_{\mu}} . \left(\sum_l \rho_{ll}^{(I)} \, \textbf{f}_{\mu,l}^{(I)} - \frac{\textbf{f}_{\mu,i}^{(I)}+\textbf{f}_{\mu,k}^{(I)}}{2}\right)
    \end{split}
\end{equation}
and $ M_\nu \ddot{\textbf{R}}_{\nu, {\rm QTSH}}^{(I)}$ represents the QTSH force of Eq.\eqref{eq:FQTSH}.

While SHXF trajectories purely follow the gradient of the active state, the forces on QTSH-XF trajectories then include not only the quantum forces of QTSH but also a new quantum-momentum-driven contribution, which is evaluated through auxiliary trajectories as in SHXF~\cite{HLM18}. This term however tends to be much smaller than the others due to the inverse nuclear mass factor, and the dependence on the product of the nuclear quantum momentum and NACV means that it likely affects more the tails of the nuclear distribution where there are less trajectories. We denote then QTSH-XF0 to represent a simpler algorithm that couples electronic coefficients via Eq. \eqref{eq:SHXFelectronic} to Eq.~(\ref{eq:FQTSH}) for the nuclear force. Comparison of results from QTSHXF and QTSHXF0 may be found in the Supplementary Material.

The terms arising from XF in the QTSH-XF algorithm formally break the energy conservation of the ensemble. However, owing to the aforementioned small contribution that these terms have, whether or not the degree of energy non-conservation in QTSH-XF is a concern in practice has to be assessed. It is worth noting that although QTSH formally achieves energy conservation when the ensemble is internally consistent~\cite{M19b}, in practice it can be violated depending on the quality of the trajectory sampling.  We will check the quality of energy conservation for both QTSH and QTSH-XF in the examples below.

We first test QTSH-XF on Tully's ECR model (see e.g. Refs~\cite{T90,VAM22} for model parameters).  Specifically, an initial gaussian wavepacket $\psi_0(R) = (2\pi \sigma^2)^{-\frac{1}{4}} \;\text{exp} \left(-\frac{(R-R_0)^2)}{4 \sigma^2}+i  k_0 x\right)$
is initialized on the lower PES at position $R_0 = -15$ a.u. with momentum $\hbar k_0 = 10$ a.u. and width $\sigma=\sqrt{2}$ a.u.
Exact results were obtained by wavepacket propagation with a timestep of $0.1$ a.u. on a grid with spacing $dx=1.25 \times 10^{-2}$ a.u. and boundaries [-35,65] a.u. QTSH-XF calculations employed 4000 trajectories Wigner-sampled according to the initial density and used a propagation timestep of $0.5$ atomic units. We compared with four other surface-hopping methods:  the original FSSH method,  the traditional energy-based decoherence method SHEDC~\cite{GP07,GPZ2010}, QTSH, and SHXF. The former two required 1000 trajectories while the latter two required 4000 trajectories for the same degree of convergence.  As in previous SHXF calculations performed on Tully's models\cite{VAM22}, the widths of gaussian densities assigned to auxiliary trajectories is set to $\sigma/10$, a tenth of the initial wavepacket's width, for QTSH-XF and SHXF simulations.

\begin{figure}[h]
\includegraphics[width=0.5\textwidth]{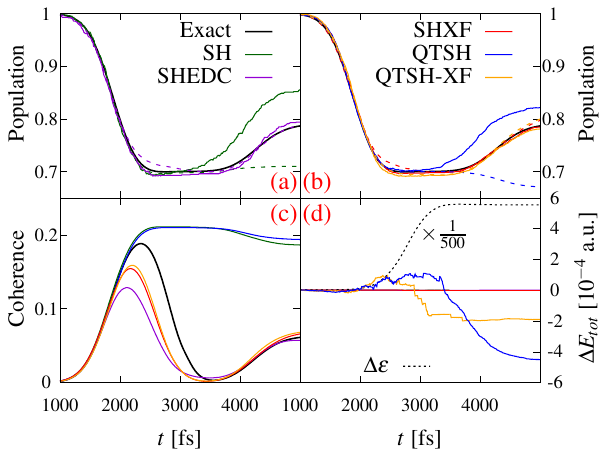}
\caption{Results for Tully's ECR model (initial conditions in the text).
a) Population ($\Pi_1$ in solid lines and $\rho_{11}$ in dashed lines) on the lower PES, predicted by the traditional methods SH in green  and SHEDC in violet along with the exact (solid black line). b) As in a) for QTSH in blue, SHXF in red, and QTSH-XF in orange.  c) Coherences for all methods compared with the exact result.  d) Error in energy conservation for all methods. The evolution of average energy difference of the PESs for QTSH-XF trajectories is also shown in black dashed line, scaled by a factor $1/500$.}
\label{fig:ECR}
\end{figure}

 The initial wavepacket has a relatively low incoming momentum, so that after the first transfer of population to the upper state in the coupling region, part of the wavepacket slides back down encountering the non-adiabatic coupling again.
This behavior is evident in the exact populations shown in Fig.~\ref{fig:ECR}, and the expected large internal inconsistency of SH (in panel (a)) and QTSH (in panel (b)) is evident after the first population transfer. This is cured by applying the first-principles decoherence term from the exact factorization: both SHXF and QTSH-XF give results close to the exact, with small internal inconsistency. In contrast to SH and QTSH which remain coherent  throughout, both SHXF and QTSH-XF also capture the coherence well (panel (c)), although decohering a little faster than the exact. 

Panel (d) in Fig.~\ref{fig:ECR} shows the error in conservation of the total energy for all methods. Note that in QTSH and QTSH-XF the energy includes a contribution from a vector potential related to the difference between the velocity and momentum in Eq.~(\ref{eq:QTSHRdot}); the formulas are given in the Supplementary Material for clarity.

The ensemble-averaged energy difference between the PESs as a function of time is also shown in black dashed lines (downscaled by a factor $1/500$).
While SH, SHEDC and SHXF perfectly conserve energy by construction, even on an individual trajectory level, QTSH and QTSH-XF display deviations. 
The violation of energy conservation in QTSH tracks the deviation from internal consistency, as expected. However, it is relatively minor here on the scale of the electronic frequencies of the system. On the other hand, QTSH-XF is much more internally consistent and its energy error is only about half as much as that of QTSH, even though it is not formally expected to conserve energy as discussed earlier. Again, on the scale of the electronic frequencies, the violation is extremely small. 
This example demonstrates that, compared with SHXF and QTSH, the  QTSH-XF scheme captures the decoherence just as well as SHXF that is missed by QTSH, and approximately conserves ensemble energy just as well or even better than QTSH and more physically than SHXF.

To see the full advantage of QTSH-XF we need to consider an example beyond one-dimension, where ambiguities from the different choices of velocity rescalings of the traditional SH and SHXF are eliminated. To this end, we next look at an eight-mode linear vibronic coupling model
of the photoexcited uracil cation~\cite{AKM15}. We will compare the same algorithms as in the previous example, with the additional consideration of velocity rescaling ansatz used (either isotropic or along NACV). 
Dynamics from the D$_2$ cationic state in this model recently demonstrated the importance of the quantum-momentum driven transitions in multi-state problems that are missing in all traditional methods (even with decoherence corrections), where exact-factorization based methods such as SHXF gave a qualitative improvement much closer to the MCTDH reference~\cite{VMM22,VVRM23}.
Since the dynamics starts from the second diabatic state in the reference MCTDH calculation~\cite{AWM16}, we do the following procedure.  For all methods,  1000 trajectories are first run starting from the adiabatic D$_0$, D$_1$ and D$_2$ with a timestep of $0.1$ fs. All observables are then obtained as weighted averages of the 3000 trajectories with respective importance of each initial state following the initial pure state of the reference MCTDH calculation: 94\% on D$_2$, 5\% on D$_1$ and 1\% on D$_0$. Gaussian width of auxilliary trajectories in SHXF and QTSH-XF are set to that of the initial wavepacket. 

\begin{figure}[h]
\includegraphics[width=.49\textwidth]{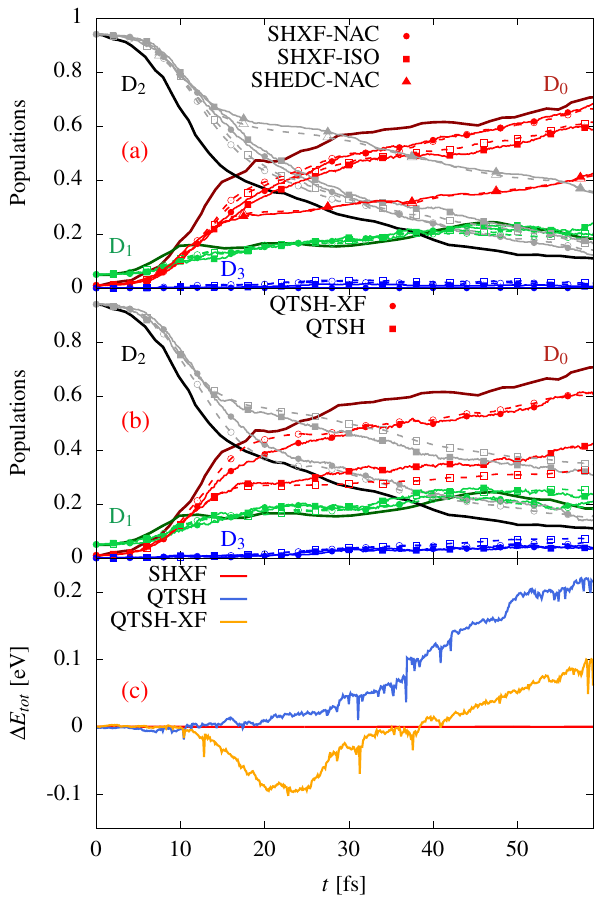}
\caption{Results for an LVC model of the uracil cation.
a) Populations  ($\Pi_i$ in solid lines and filled symbols, $\rho_{ii}$ in dashed lines and hollow symbols) for the indicated states as predicted by SHXF with velocity rescaling along the NACV and isotropic rescaling. D$_2$ and D$_0$ populations from SHEDC with scaling along NACV are shown for comparison. The reference MCTDH calculation is shown in thick lines and darker colors.
b) As in a) for the QTSH and QTSH-XF methods.
c) Error in energy conservation for QTSH, SHXF and QTSH-XF in eV. 
}
\label{fig:uracil_mixed}
\end{figure}

 The top panels of Fig.~\ref{fig:uracil_mixed} show that, while dramatically improving over the traditional SH and SHEDC in the overall trend, SHXF is noticeably sensitive to the choice of velocity rescaling. A similar sensitivity is shown by SH and SHEDC. The choice of keeping or reversing the momentum after a forbidden hop did not make so much of a difference in this case.
 
 Instead our new approach, QTSH-XF, removes this uncertainty in the prediction, and even slightly improves upon the SHXF performance at small and intermediate times. At the same time, it improves significantly upon QTSH which shows a similar behavior as the traditional SH methods. 
At longer times, QTSH-XF very slightly underperforms SHXF when the latter is performed with scaling along the NACV, but is similar to SHXF when scaling is performed isotropically. While there is a small electronic population of the D$_3$ state in SHXF (scaling along the NACV) at longer times, the fraction of trajectory measure correctly yields zero D$_3$ population, whereas QTSH-XF (and QTSH) populate D$_3$ in both measures of population shown. This may be a consequence of energy non-conservation (panel (c) of Fig.~\ref{fig:uracil_mixed}): Both QTSH-XF and QTSH see a slight increase of total energy at later times consistent with the increase of D$_3$ population, while their trend is different between 10 to 30 fs. This time interval corresponds to the maximal internal inconsistency for D$_2$ and D$_0$ populations in QTSH-XF (we can make the same comment for both SHXF calculations) and corresponds to a decrease in total energy which is compensated around 30 fs when internal consistency is retrieved. Internal inconsistency in QTSH appears slightly later and sees the opposite relation between the two measures of population compared to QTSH-XF, thus resulting in a slight rise of total energy starting at 20 fs. QTSH internal inconsistency only worsens with time. At the end of the dynamics, the total energy for QTSH has increased by 0.2 eV, while for QTSH-XF it has increased by 0.1 eV. It is interesting to note that in comparison, the coupled-trajectory CTMQC algorithm violated energy conservation by 2.2 eV while, due to numerical issues, the energy-corrected CTMQC-E algorithm was able to lower the energy increase only to 1 eV~\cite{VM23,VILMA23}. In the Supplementary Material, in addition to illustrating the very minor impact on the results of neglecting $\mathcal{F}_Q$ in Eq.~\eqref{eq:FQTSHXF}, we also investigate the effect of a more recent prescription \cite{HGM23} for QTSH equations when they are coupled with XF-based electronic evolution. Seeing no substantial improvement of our results, we favor QTSH-XF and its approximation QTSHXF0 as the original QTSH equations our algorithm is built upon are explicitly derived from the Quantum-Classical Liouville equation while the new prescription of Ref \cite{HGM23} is not.

\begin{figure}
\includegraphics[width=.48\textwidth]{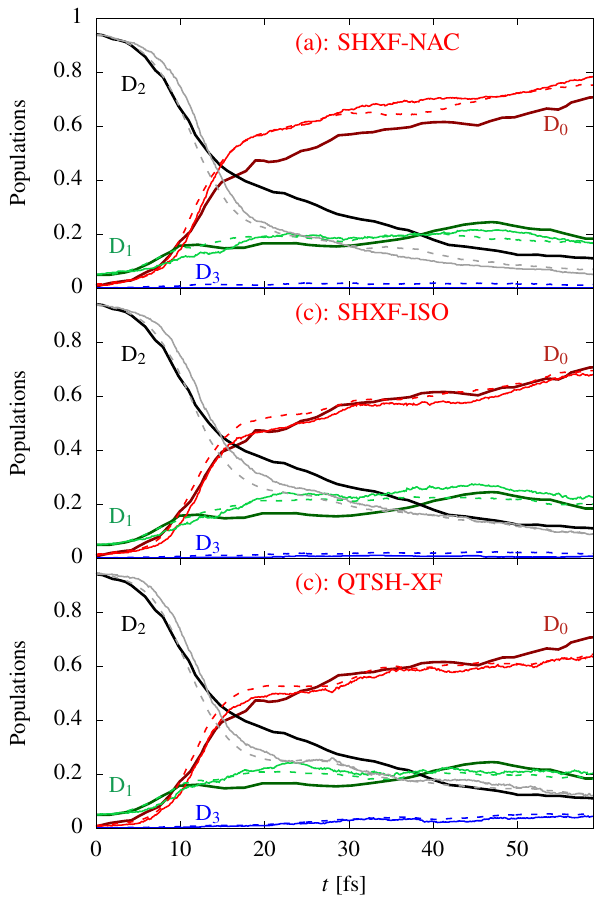}
\caption{Population dynamics in the LVC model of uracil cation with different methods  starting in the pure state (see text), shown against the MCTDH reference:
a) SHXF with velocity rescaling along NACV. b) SHXF with isotropic velocity rescaling. c) QTSH-XF.
}
\label{fig:uracil_pure}
\end{figure}

While the dynamics presented in Fig.~\ref{fig:uracil_mixed} evolve from a ``mixed state", where each trajectory is launched with an electronic population purely on the active state, the actual initial state in the reference MCTDH calculation is a pure state of the molecular wavefunction which involves a superposition of adiabatic electronic states~\cite{VMA23}. 
In Fig.~\ref{fig:uracil_pure} we approximate this pure state by associating identical real coefficients to each trajectory, with magnitude equal to the square-root of the initial adiabatic populations; the initial active states for the trajectories are distributed according to these populations. The results show that for QTSH-XF and SHXF with either scaling along NACV or isotropic, which were the three best-performing methods for the mixed state of Fig.~\ref{fig:uracil_mixed}, 
there is significantly improved agreement with the MCTDH reference for the first 5~fs, indicating that the initial conditions used here are even more faithfully capturing the MCTDH dynamics  compared to the procedure used in Fig.~\ref{fig:uracil_mixed}; further SHXF with isotropic scaling and QTSH-XF both are remarkably close to the reference MCTDH curves.  Considering SHXF,  the populations with isotropic velocity rescaling agree better with MCTDH than SHXF-NAC does (contrary to the mixed-state initial conditions); the latter overestimates the population transfer to D$_0$. QTSH-XF, although still showing D$_3$ population, closely matches the MCTDH results, except around 20~fs, where all three methods exhibit too little population in D$_2$ and too much in D$_0$.

In summary, QTSH-XF combines the best aspects of two mixed quantum-classical methods that are derived from two different, but equivalently exact, formulations of molecular quantum dynamics, while retaining computational efficiency due to its independent trajectory nature. QTSH, originating from phase space trajectories, has no need for velocity rescaling or hop rejection while still conserving energy on an ensemble level, thereby eliminating two sources of {\it ad hoc} procedures in comparison with regular SH and incorporating energy conservation in a way that is closer to the true quantum dynamics. However, QTSH suffers from overcoherence, manifesting itself in a coherent electronic evolution even away from nonadiabatic coupling regions and in the lack of internal consistency.
SHXF, on the other hand, which is derived from the exact factorization of the molecular wavefunction, 
captures decoherence from first principles due to the nuclear quantum momentum, but 
has the unsettling features of velocity rescaling and rejected hops and unphysically conserves energy on the individual trajectory level. 
The combined method QTSH-XF is of similar accuracy and efficiency as SHXF, as shown here for Tully's ECR model and an LVC model of the uracil cation. In particular, for the latter, QTSH-XF retains the qualitative improvement of SHXF over traditional methods through the quantum-momentum-driven transitions. QTSH-XF predicts a small unphysical population of a higher-lying state in the cation, which SHXF prevents in an equally unphysical way, through frustrated hops. However, QTSH-XF requires less manual input parameters and choices than SHXF, providing a less ambiguous description.
In comparison with QTSH, QTSH-XF leads to significantly improved results at the expense of introducing a new parameter, which is the 
width of the Gaussian sitting at the auxiliary trajectories' positions mimicking a contribution to the nuclear density (which is needed to calculate the quantum momentum). While the additional terms do not formally guarantee the conservation of the ensemble energy, in practice the QTSH-XF energy varies less than in QTSH, which is most likely due to the much improved internal consistency.

\acknowledgements{
Financial support from the National Science Foundation Award CHE-2154829 (NTM, AR),  the Computational Chemistry
Center: Chemistry in Solution and at Interfaces funded by the
U.S. Department of Energy, Office of Science Basic Energy
Sciences, under Award DE-SC0019394 as
part of the Computational Chemical Sciences Program (LD), and the German Academic Scholarship Foundation (AR) are gratefully acknowledged.}

\bibliography{ref_na.bib}

\newpage
\clearpage

\appendix

\onecolumngrid

\subsection{\large Supplementary Material}

\subsection{I. Total energy expression in QTSH}
The total energy is the expectation value of the total Hamiltonian, $\hat{H} = \hat{H}_{\rm BO} + \hat{T}_n$ where $\hat{T}_n$ is the nuclear kinetic energy, which, for MQC methods becomes the trajectory sum
\ben
E = \frac{1}{N_{tr}}\sum_J\left(\frac{1}{2}\sum_\nu M_\nu^{(J)} \dot{R}_\nu^{(J)2} + U^{(J)}\right)
\label{eq:Ecl}
\een
where $U^{(J)}$ denotes the electronic energy. For SH methods,  $U^{(J)} = \eps_a^{(J)}$, the BO potential energy of the active state, rather than being defined through the weighted sum defined by the electronic populations. The total energy in SH is conserved~\cite{VILMA23} because the force on the trajectories is conservative while the trajectory remains on state $a$,  and when a hop occurs to a different active state, velocity-adjustment is then imposed to conserve energy. 
In QTSH, Eq.~\ref{eq:Ecl} involves the kinematic momentum $M_\nu\dot R_\nu$ which differs from the canonical momentum $P_\nu$ (Eq. 6a of the main text), and writing the energy in terms of $P_\nu$ yields
\ben\label{eq:QTSH_E}
    E_{\text{QTSH}} = \frac{1}{N_{tr}}\sum_J\left(\sum_\mu\frac{\textbf{P}^{(J)2}_\mu}{2 M_\mu} + U^{(J)} - 2\hbar \sum_{i<j} \sum_\mu \frac{\mathbf{d}_{\mu,ij}^{(J)} \cdot \mathbf{P}^{(J)}_\mu}{M_\mu} \, \text{Im} \Big\{ \rho^{(J)}_{ij} \Big\} \right)
\een
neglecting again terms of order $(\hbar/M_{\mu})^2$.
The third term is referred to as coherence energy in Ref.~\cite{M19b}\footnote{Note that a factor of $1/M_\mu$ is missing in the original publication, Ref.~\cite{M19b}}, partly because of its off-diagonal character in the density-matrix. Eq.~\ref{eq:QTSH_E} is equivalent to the trace $Tr[\rho H]$, which is how it is obtained in Ref.~\cite{M19b}.
No velocity adjustment is made when a hop to another active state occurs, thus the energy of an individual trajectory is not constrained to be conserved. In all results presented in the main text and supplementary material,  Eq.~\eqref{eq:QTSH_E} was used to compute the total energy of QTSH and QTSH-XF while Eq.~\eqref{eq:Ecl} was used for other methods.

\subsection{II. Performance of QTSH-XF variants}
The derivation of QTSH-XF in the main text introduced new elements whose relative impact on the dynamics we study here. 

\subsubsection{1. QTSH-XF0}

\begin{figure}[!htb]
    \centering
    \includegraphics[width=.65\textwidth]{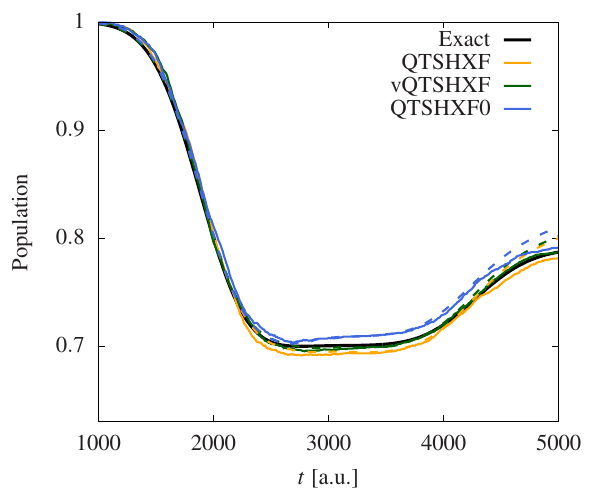}
    \caption{Ground state population ($\Pi_i$ in solid lines and filled symbols, $\rho_{ii}$ in dashed lines and hollow symbols) for the ECR model (see main text for system definition): Comparison of QTSH-XF to QTSH-XF0. Exact wavepacket results are shown in black solid lines.}
    \label{fig:pop_ECR_QTXF0}
\end{figure}

\begin{figure}[!htb]
    \centering
    \includegraphics[width=.7\textwidth]{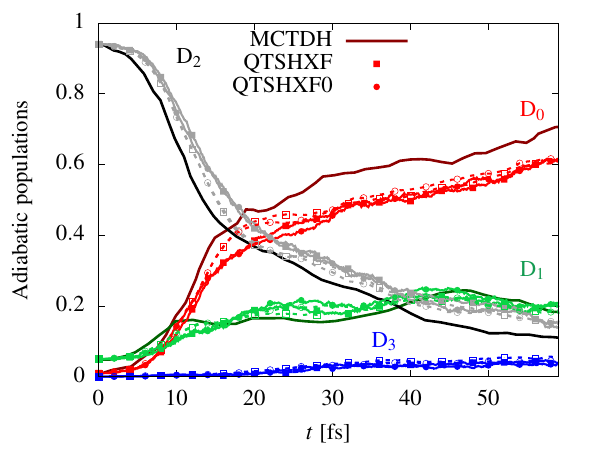}
    \caption{Population dynamics ($\Pi_i$ in solid lines and filled symbols, $\rho_{ii}$ in dashed lines and hollow symbols) for the LVC model of the uracil cation: Comparison of QTSH-XF to QTSH-XF0. MCTDH results shown with darker and thicker solid lines. Trajectories are initialized as a mixed state (see main text).}
    \label{fig:pops_Uracil_QTXF0}
\end{figure}

\begin{figure}[!htb]
    \centering
    \includegraphics[width=.58\textwidth]{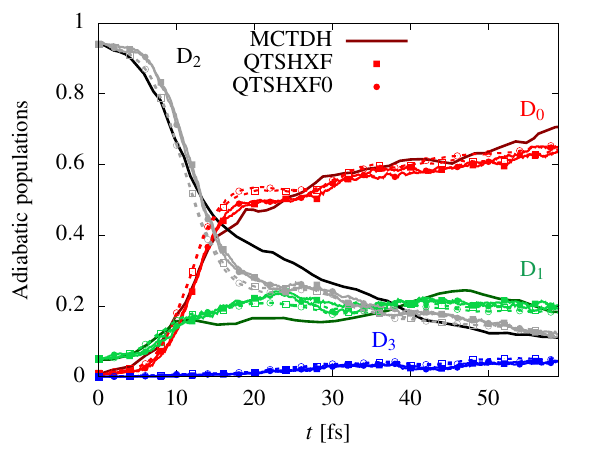}
    \caption{Population dynamics ($\Pi_i$ in solid lines and filled symbols, $\rho_{ii}$ in dashed lines and hollow symbols) for the LVC model of the uracil cation: Comparison of QTSH-XF to QTSH-XF0. MCTDH results shown with darker and thicker solid lines. Trajectories are initialized as a pure state (see main text).}
    \label{fig:pops_Uracildens_QTXF0}
\end{figure}

\begin{figure}[!htb]
    \centering
    \includegraphics[width=.7\textwidth]{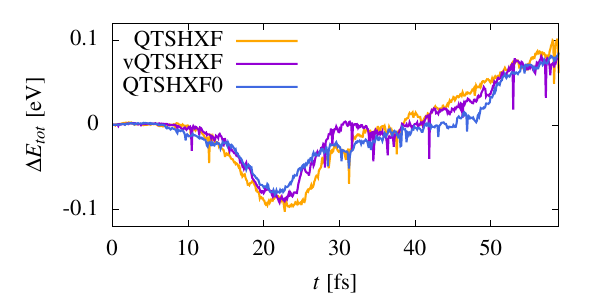}
    \caption{Error in energy conservation for QTSHXF and vQTSHXF in percentage of the total energy for the LVC model of the uracil cation. Trajectories are initialized as a mixed state (see main text).}
    \label{fig:E_Uracil_QTXF0}
\end{figure}

First, the QTSH-XF force [see Eq. (8) in the main text] introduces an additional quantum-momentum-driven force on the nuclei on top of the QTSH forces. We argued in the main text that this is expected to be small, and defined the QTSH-XF0 algorithm in which only the QTSH force is kept, coupled with the XF electronic equation. Here we show the difference explicitly on the population dynamics of the ECR model (Fig. \ref{fig:pop_ECR_QTXF0}) and the uracil cation model (Figs. \ref{fig:pops_Uracil_QTXF0} and \ref{fig:pops_Uracildens_QTXF0} for mixed and pure state initialization, respectively). Numerical parameters used are the same as for calculations in the main text. It is clear that the results are not strongly affected in either case. We compare the quality of energy conservation with and without this term in Fig. \ref{fig:E_Uracil_QTXF0}, again finding a close agreement between QTSH-XF0 and QTSH-XF.

Owing to the minor impact of the approximation QTSH-XF0, this study suggests that the term involving the quantum momentum in the QTSH-XF force does not play an important role in capturing non-adiabatic dynamics faithfully, nor does it constitute a sizeable source of energy non-conservation in QTSH-XF.     

\clearpage

\subsubsection{2. vQTSH-XF}

Next we compare the use of the original prescription of QTSH\cite{M19b} to a more recent suggestion in Ref.\cite{HGM23} by its original proponent in the making of QTSH-XF. We will call this alternate version of our algorithm vQTSH-XF and clarify the differences with QTSH-XF below. 

In the most recently proposed algorithm for QTSH (which we will refer to as vQTSH), the electronic evolution as well as the hopping probability are obtained as in fewest-switches SH, whereas QTSH replaces the velocity $\dot{\textbf{R}}_\nu^{(I)}$ by $\textbf{P}_\nu^{(I)}/M_\nu$. This leads to the following equations for vQTSH-XF, compared line-by-line with QTSH-XF:

\begin{equation}
\def\arraystretch{2.5}
\setlength{\tabcolsep}{12pt}
\begin{tabular}{cc}
     $\text{QTSH-XF}$  & $\text{vQTSH-XF}$ \\
    \hline
     $\displaystyle\dot C_l^{(I)} = -\frac{i}{\hbar}\epsilon^{\mathrm{BO}}_l(\textbf{R}^{(I)}) C_l^{(I)} - \sum_k C_k^{(I)} \sum_{\nu} \frac{\textbf{P}_\nu^{(I)}}{M_\nu}\cdot\textbf{d}_{\nu,lk}^{(I)}$  &  $\displaystyle\dot C_l^{(I)} = -\frac{i}{\hbar}\epsilon^{\mathrm{BO}}_l(\textbf{R}^{(I)}) C_l^{(I)} - \sum_k C_k^{(I)} \sum_{\nu} \dot{\textbf{R}}_\nu^{(I)}\cdot\textbf{d}_{\nu,lk}^{(I)}$\\
    $ + \; \dot C_{l,\text{XF}}^{(I)}$     &  $+ \; \dot C_{l,\text{XF}}^{(I)}$  
    \\
    &
    \\
    $\displaystyle P_{a\rightarrow k}^{(I)} = {\rm max} \left\{0,-\frac{\displaystyle 2 \text{Re} \{C^{(I)*}_a C^{(I)}_k\} \sum_\nu \textbf{d}_{\nu,lk}^{(I)}\cdot \frac{\textbf{P}_\nu^{(I)}}{M_\nu}}{|C^{(I)}_a|^2} \Delta t \right\}$ & $\displaystyle P_{a\rightarrow k}^{(I)} = {\rm max} \left\{0,-\frac{\displaystyle 2 \text{Re} \{C^{(I)*}_a C^{(I)}_k\} \sum_\nu \textbf{d}_{\nu,lk}^{(I)}\cdot \dot{\textbf{R}}^{(I)}_\nu}{|C^{(I)}_a|^2} \Delta t \right\}$
\end{tabular}
\end{equation}
Moreover, the equations of motion for trajectories in vQTSH read:

\begin{subequations}
\begin{align} \label{eq:vQTSHRdot}
    \dot{\textbf{R}}^{(I)}_\nu =& \frac{\textbf{P}^{(I)}_\nu}{M_\nu} - \frac{2\hbar}{M_\nu} \sum_{i<j} \text{Im}\big\{\rho_{ij}^{(I)}\big\} \textbf{d}^{(I)}_{\nu,ij} \\
    \begin{split}
    \dot{\textbf{P}}_\nu^{(I)} =& - \nabla_\nu \epsilon^\mathrm{BO}_a(\textbf{R}^{(I)}) \\
    &+ 2\hbar \sum_\mu\sum_{i<j}   \text{Im}\{\rho_{ij}^{(I)}\}  (\dot{\textbf{R}}^{(I)}_\mu \cdot \nabla_\mu )  \textbf{d}_{\nu,ij}^{(I)} \label{eq:vQTSHRdot2}
    \end{split}
\end{align}
\end{subequations}
which differs from QTSH in the last line.
Combining these equations and using $(\dot{\textbf{R}}^{(I)}_\mu \cdot \nabla_\mu )  \textbf{d}_{\nu,ij}^{(I)} = \dot{\textbf{d}}_{\nu,ij}^{(I)}$ one finds the force used in vQTSH-XF, that we give side-to-side with QTSH-XF force:

\begin{equation}
\def\arraystretch{2.5}
\setlength{\tabcolsep}{12pt}
\begin{tabular}{cc}
     $\text{QTSH-XF}$  & $\text{vQTSH-XF}$ \\
    \hline
     $\displaystyle M_\nu \ddot{\textbf{R}}^{(I)}_{\nu} = -\nabla_\nu \epsilon^\mathrm{BO}_a(\textbf{R}^{(I)})
        + 2 \sum_{i<j}\Delta \epsilon^{\rm BO}_{ij} \text{Re}\{\rho_{ij}^{(I)}\} \textbf{d}^{(I)}_{\nu,ij}$  &  $\displaystyle M_\nu \ddot{\textbf{R}}^{(I)}_{\nu} = -\nabla_\nu \epsilon^\mathrm{BO}_a(\textbf{R}^{(I)})
        + 2 \sum_{i<j}\Delta \epsilon^{\rm BO}_{ij} \text{Re}\{\rho_{ij}^{(I)}\} \textbf{d}^{(I)}_{\nu,ij}$\\
    $\displaystyle + 2 \hbar \sum_{i<j}  \textbf{d}^{(I)}_{\nu,ij} \sum_{l}\sum_{\mu} \frac{\textbf{P}^{(I)}_{\nu}}{M_\nu} \cdot \text{Im}\left\{\textbf{d}_{\mu,jl}^{(I)} \rho_{il}^{(I)} -\textbf{d}_{\mu,il}^{(I)}\rho_{jl}^{(I)} \right\}$     &  $\displaystyle + 2 \hbar \sum_{i<j}  \textbf{d}^{(I)}_{\nu,ij} \sum_{l}\sum_{\mu} \dot{\textbf{R}}^{(I)}_{\nu} \cdot \text{Im}\left\{\textbf{d}_{\mu,jl}^{(I)} \rho_{il}^{(I)} -\textbf{d}_{\mu,il}^{(I)}\rho_{jl}^{(I)} \right\}$  
    \\
    $+\mathcal{F}_Q$ & $+\mathcal{F}_Q$
\end{tabular}
\end{equation}

We recall that in order to obtain the expression for $M_\nu \ddot{\textbf{R}}^{(I)}_{\nu, {\rm QTSH-XF}}$ we neglected a $\hbar^2/M_\mu$ term related to $([M_\mu\dot{\textbf{R}}^{(I)}_\mu - \textbf{P}^{(I)}_\mu] \cdot \nabla_\mu )  \textbf{d}_{\nu,ij}^{(I)}$. One of the appeals of vQTSH is that no such term is to be neglected here, making it potentially better at conserving the total energy. The definition of vQTSH-XF total energy is the same as vQTSH:

\ben\label{eq:vQTSH_E}
    E_{\text{vQTSH}} = \frac{1}{N_{tr}}\sum_J\left(\sum_\mu\frac{\textbf{P}^{(J)2}_\mu}{2 M_\mu} + U^{(J)} - 2\hbar \sum_{i<j} \sum_\mu \mathbf{d}_{\mu,ij}^{(J)} \cdot \dot{\textbf{R}}^{(I)}_{\mu} \, \text{Im} \Big\{ \rho^{(J)}_{ij} \Big\} \right)
\een

\begin{figure}[!htb]
    \centering
    \includegraphics[width=.68\textwidth]{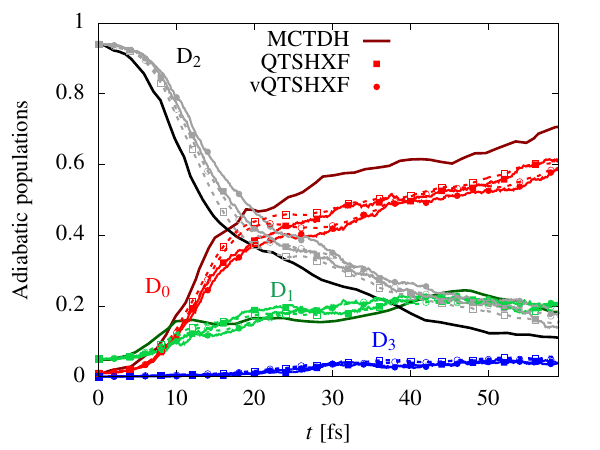}
    \caption{Population dynamics ($\Pi_i$ in solid lines and filled symbols, $\rho_{ii}$ in dashed lines and hollow symbols) for the LVC model of the uracil cation: Comparison of QTSH-XF to QTSH-XF0. MCTDH results shown with darker and thicker solid lines. Trajectories are initialized as a mixed state (see main text).}
    \label{fig:pops_Uracil_vQTXF}
\end{figure}

\begin{figure}[!htb]
    \centering
    \includegraphics[width=.68\textwidth]{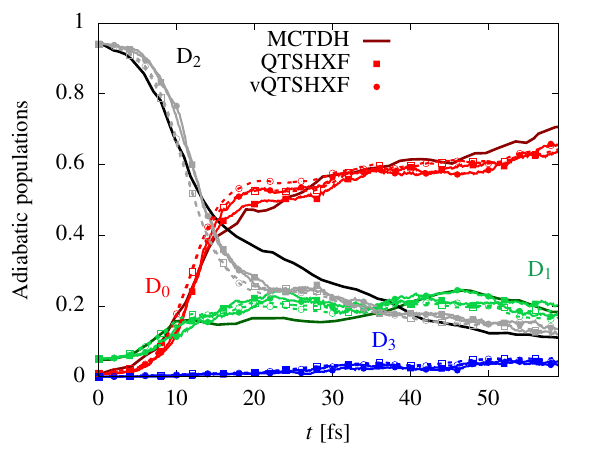}
    \caption{Population dynamics ($\Pi_i$ in solid lines and filled symbols, $\rho_{ii}$ in dashed lines and hollow symbols) for the LVC model of the uracil cation: Comparison of QTSH-XF to QTSH-XF0. MCTDH results shown with darker and thicker solid lines. Trajectories are initialized as a pure state (see main text).}
    \label{fig:pops_Uracildens_vQTXF}
\end{figure}

Figs \ref{fig:pops_Uracil_vQTXF} and \ref{fig:pops_Uracildens_vQTXF} show the population dynamics of the uracil cation for mixed and pure state initialization respectively, and it is seen that it only modestly affects the results. A similar comparison for the ECR model in Fig \ref{fig:pop_ECR_QTXF0} yields the same conclusion. Moreover, a comparison of energy change as a function of time as done on Fig \ref{fig:E_Uracil_QTXF0} shows it does not lead to an appreciable improvement in energy conservation.

Seing that the performance of the two algorithms is so similar, we are lead to favor QTSH-XF over vQTSH-XF as the original equations of QTSH have been derived from first principles.

\end{document}